\pdfoutput=1

\documentclass[3p,times,twocolumn]{elsarticle}

\usepackage{ecrc}

\volume{00}

\firstpage{1}

\journalname{Nuclear Physics B Proceedings Supplement}

\runauth{Konstantin A. Kouzakov, Alexander I. Studenikin, and
Mikhail B. Voloshin}

\jid{nuphbp}

\jnltitlelogo{Nuclear Physics B Proceedings Supplement}

\usepackage{amssymb}
\usepackage{bm}
\usepackage{graphicx}

\usepackage[figuresright]{rotating}

\begin{document}

\begin{frontmatter}



\dochead{}

\title{On neutrino-atom scattering in searches for neutrino magnetic moments}


%

\author[KAK]{Konstantin A. Kouzakov}
\author[AIS1,AIS2]{Alexander I. Studenikin\corref{cor1}}
\ead{studenik@srd.sinp.msu.ru}
\author[MBV1,MBV2]{Mikhail B. Voloshin}

\address[KAK]{Department of Nuclear Physics and Quantum Theory of Collisions, Faculty of Physics, Moscow State University, Moscow 119991, Russia}
\address[AIS1]{Department of Theoretical Physics, Faculty of Physics, Moscow State University, Moscow 119991, Russia}
\address[AIS2]{Joint Institute for Nuclear Research, Dubna 141980, Moscow Region, Russia}
\address[MBV1]{William I. Fine Theoretical Physics Institute, University of Minnesota, Minneapolis, Minnesota 55455, USA}
\address[MBV2]{Institute of Theoretical and Experimental Physics, Moscow, 117218, Russia} \cortext[cor1]{Corresponding author}

\begin{abstract}
In the experimental searches for neutrino magnetic moments using
germanium detectors one studies the ionization channel in the
neutrino-atom scattering. We find that the so-called stepping
approximation to the neutrino-impact ionization is exact in the
semiclassical limit, and that the deviations from this
approximation are very small.
%
\end{abstract}

%

\end{frontmatter}


%
%
The neutrino magnetic moments (NMM) expected in the Standard Model
are very small and proportional to the neutrino masses: $\mu_\nu
\approx 3 \times 10^{-19}\, \mu_B \, (m_\nu/1 \, eV)$, with $\mu_B
= e/2m_e$ being the electron Bohr magneton, and $m_e$ is the
electron mass. Any larger value of $\mu_\nu$ can arise only from
physics beyond the Standard Model~\cite{giunti09}. Current direct
experimental searches for a magnetic moment of the electron
(anti)neutrinos from reactors have lowered the upper limit on
$\mu_\nu$ down to $\mu_\nu < 3.2 \times 10^{-11} \,
\mu_B$~\cite{gemma10}. At small energy transfer $T$ the inclusive
cross section for the magnetic neutrino scattering on a free
electron behaves as $d\sigma_{(\mu)}/dT\propto1/T$~\cite{vogel89},
while that due to weak interaction, $d\sigma_{(w)}/dT$, is
practically constant in $T$~\cite{vogel89}. The current
experiments using Ge detectors have reached threshold values of
$T$ as low as few keV, where one can expect a modification of the
free-electron formulas due to the binding of electrons in the Ge
atoms. In a recent paper~\cite{wong10}, a significant enhancement
of the NMM contribution by the atomic ionization effects was
claimed. And later on, the authors of Ref.~\cite{wong10} disproved
their claim (see also Refs.~\cite{voloshin10,ks}). Our recent
theoretical analysis~\cite{ksv}, involving the WKB and
Thomas-Fermi models and accounting for electronic correlations,
has shown that the so-called stepping approximation (SA),
originally introduced in Ref.~\cite{kmsf} from an interpretation
of numerical data, works with a very good accuracy. SA treats the
process as scattering on independent electrons occupying atomic
orbitals and suggests that the cross sections follow the
free-electron behaviors down to $T$ equal to the ionization
threshold for the orbital; and below that energy the electron on
the corresponding orbital is `inactive' thus producing a sharp
`step' in the dependence of the cross section on $T$. We thus
argue that SA can be applied to the analysis of the present and
future data of searches for NMM with Ge detectors down to the
values of the energy deposition $T \sim 0.3$~keV.

One of the authors (A.I.S.) thanks Prof. Tzanakos for the
invitation to attend the Neutrino 2010 Conference.







%

%





\end{document}